# SIMULATIONS OF THE CHAIN LENGTH DEPENDENCE OF THE MELTING MECHANISM IN SHORT-CHAIN N-ALKANE MONOLAYERS ON GRAPHITE

Cary L. Pint

Department of Physics, University of Northern Iowa, Cedar Falls, IA  50614

Electronic Mail:  cpint@uni.edu

## ABSTRACT

The melting transition in solid monolayers of a series of short-chained *n*-alkanes, *n*-octane (*n*-$C_8H_{18}$), *n*-decane (*n*-$C_{10}H_{22}$), and *n*-dodecane (*n*-$C_{12}H_{26}$) physisorbed onto the graphite basal plane are studied through use of molecular dynamics simulations.  Utilizing previous experimental observations of the solid phase behavior of these monolayers, this study investigates the temperature dependence of the phases and phase transitions in these three monolayers during the solid-fluid phase transition, and compares the observed melting behavior to previous studies of hexane and butane monolayers.  In particular, this study seems to indicate a greater dependence of the melting transition on the formation of gauche defects in the alkyl chains as the chain length is increased.  In light of the previously proposed "footprint reduction" mechanism and variations where the formation of gauche defects are energetically negated, simulations seem to suggest that decane and dodecane monolayers are generally equally as dependent upon the formation of gauche defects for the melting transition to take place, whereas octane monolayers seem to have less dependence, but follow a trend that is established in previous studies of melting in butane and hexane monolayers.  Also, the phase transition from a solid herringbone phase into an orientationally ordered "intermediate" phase is found to exhibit some differences as compared to a recent study of hexane monolayers, which may be interpreted as originating from the greater influence of gauche defects. Comparison to experimental melting temperatures is provided where possible, and applications involving thin film manipulation and lubrication is discussed.

**Keywords**: Molecular Dynamics, *n*-alkanes, adsorption, graphite, octane, decane, dodecane

# 1. INTRODUCTION

The study of physisorbed systems contributes to many fields that are currently expanding due to the increased current attention focused upon systems that exist on very small length scales. With the explosion in technology over the past few decades that allows the study of nanoscale systems experimentally and through the use of computer simulations, the understanding of many phenomena that exists at such scales is being studied for many practical and industrial applications. In particular, computer simulations play an important role in these studies, as many experimental studies are very expensive and time-consuming to conduct. Thus, simulations, properly coupled with existing experimental work, can be used to explore many situations that have not yet been the primary focus of experimental study, or else explore aspects of systems that have been studied experimentally to probe the phase behavior at a deeper level.

In this study, monolayers of a series of short-chain alkanes are studied, physisorbed onto the graphite basal plane. Although there are many species of interesting molecular adsorbates, the *n*-alkanes have received an ample amount of experimental and theoretical study in recent years because of their simple straight-chained structure, and high potential for useful applications (since *n*-alkanes contribute to many surface problems dealing with topics such as lubrication, adhesion, detergency, etc.). The *n*-alkanes have also been proposed as a simple model [1,2] for more complex adsorbed molecules, whose study through simulation and experiment is often difficult, due to the complex molecular structures and the extremely long time scales needed to probe intramolecular motion that cannot be readily achieved with MD simulations at this point. Therefore, not only are the *n*-alkanes useful in their own right, but they are also extremely important for understanding molecular behavior of many other types of complex molecules that are still posing significant theoretical questions today. Graphite has been a popular choice as an adsorbate in both experiment and computation alike, due to the vast availability and the symmetry that graphite exhibits as a substrate.

The three alkanes studied in this work, octane, decane, and dodecane (henceforth referred to as C8, C10, and C12 respectively), have been subjects of recent experimental study, but to the knowledge of the author have not been the subject of a study focusing on the phase behavior and/or phase transitions as of yet. Although one may observe that these are three specific molecules that are a larger part of a vast family of molecules with similar properties, the increasing chain length of these molecules as compared to previously studied monolayers of hexane and butane involves more degrees of freedom that allow more complex molecular behavior, which could serve to emphasize the dependence of the phases and phase transitions on this behavior.

There has been a notable amount of recent experimental work [1-14] that studies the phase transitions and structure of short and intermediate-chained *n*-alkanes on graphite. Specifically for the adsorbed systems dealt with in this study[5-9], there has been work that utilizes X-ray and neutron diffraction [3,5,6] differential scanning calorimetry (DSC) [3,4], and nuclear magnetic resonance (NMR) [7] to study these systems physisorbed at submonolayer, monolayer, and multilayer coverages onto graphite. These studies find that the even-numbered alkanes [5] (as studied in this work) arrange in an isomorphic HB phase at low temperatures (with $6 \leq n \leq 10$) that is commensurate with the substrate. This work also finds that fully commensurate monolayers of C12 do not arrange in a HB phase, but rather exhibit a low-temperature rectangular-centered (RC) phase that is commensurate with the graphite substrate and generally resembles the phases observed in both odd alkane [8] and longer-chained [3,4] alkane monolayers. A DSC study [4] of monolayer C12 indicates that melting occurs at 287.8K from an RC phase structure, and diffraction studies for C8 and C10 indicate that there are stable monolayers that exist above the bulk melting points of 217K and 243K respectively [5].

There have also been experimental [8-14] and theoretical [13-18] studies of the melting transition and phase behavior in other *n*-alkanes, with a concentration of work existing over monolayers of hexane (C6). [11-18] The first works over C6 [13,14] suggest a footprint mechanism that requires a type of space reduction to preempt the melting transition and that melting (in C6 monolayers) occurs via gauche defects. Later work including revisions [16] to the simulated surface interaction indicates that C6 and C4 both melt through an out-of-plane orientational tilting mechanism, as opposed to gauche defects, however, this system still obeys the proposed footprint reduction mechanism in previous work. Most recent work [18] over hexane monolayers has found that there is a small contribution from gauche defects that contribute to the melting transition, but the primary mechanism for melting in C6 is tilting of the molecules, which in turn creates a space reduction that preempts the phase transition. Variations are performed which indicate that the melting temperature of the hexane monolayer is augmented by ca. 20K if gauche defect formation is eliminated.

Further experimental and theoretical studies of monolayers and multilayers of *n*-alkanes with longer chain lengths (C24 and C32) [1,2,8-10] has found that melting in these systems takes place primarily via formation of a large number of molecules with a significant number of gauche defects that contribute to the space reduction theory that was proposed for shorter-chained *n*-alkanes. As opposed to observations in shorter chained *n*-alkanes, there are no effects of tilting behavior reported, and the melting mechanism takes place completely through the formation and presence of gauche defects that mediate the solid-fluid phase transition.

Therefore, at this point in time, simulations and experiment that have studied solid physisorbed *n*-alkane monolayers gives two seemingly different pictures of the same melting transition for alkanes with different chain lengths. Interestingly, recent studies involving molecular dynamics simulations [19,20] and temperature programmed desorption (TPD) [21,22] of *desorption* of *n*-alkanes from solid surfaces also seems to indicate some anomaly in the desorption behavior with increasing chain length. In fact, Fichthorn et al. [19,20] discuss the possibility of the increasing chain length affecting the nonlinear increase of desorption energy with increasing chain length that is observed in experiment. [21,22] The presence of an increased number of molecules in a gauche conformation for longer alkanes is proposed to have an effect on desorption and thus it is relevant to wonder what further effect on physisorbed monolayers that more availability for a molecule to be in a gauche conformation presents. In particular, the scientific motivation of this work is to use simulations to study how the phase transition behavior in short-chained even *n*-alkanes evolves with increasing chain length. Therefore, the one question whose answer is sought in this study is "How does the melting transition mechanism evolve as the chain length (and hence, the more possibilities for the molecule to adopt a gauche conformation) increases, and what chain length does the footprint mechanism adopt terminal gauche defect dependence."

## II. SIMULATION DETAILS

### A. Methods

To perform simulations of C8, C10, and C12, physisorbed onto the basal plane of graphite, a constant ($N$, $\rho$, $T$) canonical ensemble molecular dynamics method is utilized with a velocity Verlet RATTLE [23] algorithm to perform integration of the equations of motion. As is the case in most studies of this type, a model in which the complexity of the alkane molecule is reduced to a number of psuedoatoms, each representing the methyl ($CH_3$) and methylene ($CH_2$) groups is utilized. Due to the computation of atomic force interactions, it is often preferred to use such a model. The model most predominantly used in this study is the united-atom (UA) [24] model, which has been the primary model used in previous simulations of these types of systems. In this model, each methyl ($CH_3$) and methylene ($CH_2$) group is considered as a psuedoatom that is distinguished only by mass in the simulation. This model has been used in previous work that studies C6 on graphite, [13-18] and is found to be in good agreement with experiment. It has also been used in simulations of larger molecules, such as C24 and C32, [1,2] where it has also has given phase behavior in excellent agreement with experimental predictions and observations.

The second model is a version of the anisotropic united-atom (AUA4) model [25] whose interaction parameters have been optimized to give better agreement to experimentally observed equilibrium properties of *n*-alkanes. The difference between this model and the UA model is that in this case, the psuedoatoms are not only distinguished by mass, but have carbon atom centers shifted slightly toward the hydrogen atom positions to give a slightly more realistic picture of the molecule. Also, for simulations that utilize the AUA4 model, a different version of the dihedral potential (as will be described later), as proposed by Toxvaerd [26] is utilized in order to provide support to that used to present the results of this work. Although this model will not be utilized to explicitly report the results, comments will be made at the conclusion of this paper regarding how the results from the AUA4 model compare to the UA model. In previous work, Peters and Tildesley report [16] that the AUA model predicts a smaller number of molecules in the gauche conformation as opposed to the UA model, but that the phase behavior and transitions are generally quite similar. Also, one expects that the major contributions from the distinguished interaction of the methyl and methylene groups would diminish as the chain length of the alkane molecule increases.

All simulations that are carried out are started from an initial configuration that is representative of the monolayer commensurate phase, as studied through diffraction and reported in [5]. To obtain a spectrum of temperature points, intervals of 2-5K near transitions and 10K far away from transitions are used to classify both the solid and liquid phase behavior. To maintain temperature control, a method of velocity rescaling is used [18] which rescales the velocities to satisfy equipartition for the center-of-mass, rotational, and internal temperatures.

The unit cell parameters, computational cell sizes, simulated temperature ranges, and the total numbers of pseudoatoms in each simulation are presented in table I. On average, each simulation is carried out for a period of 700 ps (each 1 ps is 1000 simulation time steps), with the first 200 ps serving to equilibrate the system from the initial configuration. Analysis of system energies and phase behavior of the monolayer after the 200 ps equilibration period indicates that this equilibration period is sufficient to achieve a thermodynamic equilibrium. Periodic boundary conditions are imposed on the plane of the substrate, and the region above the substrate is modeled as being infinite in extent.

The time step used in all simulations is 1 fs, which is sufficient to capture the intramolecular motions that occur at the temperatures and simulation conditions in this study. Since the fast C-C stretching modes are, to a good approximation, fixed at their equilibrium values in such conditions, the RATTLE [23] algorithm is used to constrain the bond lengths to a fixed value of 1.54Å for the UA model and 1.535Å for the AUA4 model.

### B. Potential Model

In all simulations, the model that is used consists of both non-bonded and bonded interactions. The non-bonded interactions consist of (*i*) the adsorbate-adsorbate interaction, modeled by a Lennard-Jones 12-6 pair potential function given by:

$$u_{LJ}(r_{ij}) = 4\varepsilon_{ij}\left[\left(\frac{\sigma_{ij}}{r_{ij}}\right)^{12} - \left(\frac{\sigma_{ij}}{r_{ij}}\right)^{6}\right], \quad (1)$$

where $\varepsilon$ refers to the well depth of the potential, and $\sigma$ represents the collision diameter. Lorentz-Berthelot combining rules

$$\sigma_{ij} = \frac{\sigma_i + \sigma_j}{2}, \quad \varepsilon_{ij} = \sqrt{\varepsilon_i \varepsilon_j}, \quad (2)$$

describe mixed interactions when the two interacting adatoms are of different types, and (*ii*) the adsorbate-substrate interaction, which is modeled by a Fourier expansion proposed by W.A. Steele [27] and is of the form:

$$u_i^{gr} = E_{0i}(z_i) + \sum_{n=1}^{\infty} E_{ni}(z_i) f_{ni}(x_i, y_i). \quad (3)$$

In eqn (3), there are two energy terms that are derived from the pair interaction of an adatom with a smooth graphite substrate. The holding potential, defined as

$$E_{0i}(z_i) = \frac{2\pi q \varepsilon_{gr} \sigma_{gr}^6}{a_s}\left(\frac{2\sigma_{gr}^6}{45d(z_i+0.72d)^9} + \frac{2\sigma_{gr}^6}{5z_i^{10}} - \frac{1}{z_i^4} - \frac{2z_i^2 + 7z_i d + 7d^2}{6d(z_i+d)^5}\right), \quad (4)$$

is representative of the vertical forces on an adatom with respect to the substrate. In eqn 4, d corresponds to the distance between the graphene layers, $a_s$ is the surface area of a graphite unit cell, and $q$ is the number of carbon atoms in a graphite unit cell. In addition, the surface corrugation is represented by the second term in eqn (3),

$$E_{ni}(z_i) = \frac{2\pi \varepsilon_{gr} \sigma_{gr}^6}{a_s}\left[\left(\frac{\sigma_{gr}^6}{30}\right)\left(\frac{g_n}{2z_i}\right)^5 K_5(g_n z_i) - 2\left(\frac{g_n}{2z_i}\right)^2 K_2(g_n z_i)\right], \quad (5)$$

and

$$f_1(x_i, y_i) = -2\cos\left[\frac{2\pi}{a}(x+\frac{y}{\sqrt{3}})\right] - 2\cos\left[\frac{2\pi}{a}(x-\frac{y}{\sqrt{3}})\right] - 2\cos\left[\frac{4\pi}{a}(\frac{y}{\sqrt{3}})\right]. \quad (6)$$

where the $K$'s are modified Bessel functions of the second kind, and $g_n$ is the modulus of the $n^{th}$ graphite reciprocal lattice vector. This term presents the lateral forces on an adsorbate atom, and is defined only for $n=1$, as the corrugation from the subsequent graphene layers is negligible.

The bonded interactions in this system are composed of two types: bond angle bending, and dihedral angle bending. In all simulations, the bond-angle potential is given by Martin and Siepmann [24],

$$u_{bend} = \frac{1}{2}k_\theta(\theta_b - \theta_0)^2, \quad (7)$$

where $\theta_b$ is the bond angle, $\theta_0$ is the equilibrium bond angle and $k_\theta$ is the angular stiffness. For the UA model, the dihedral bending (torsional) potential is of the form [28]:

$$u_{tors} = \sum_{i=0}^{5} c_i (\cos\phi_d)^i, \quad (8)$$

and for the AUA4 model, the potential function is of the same exact form, except the sum is over a series of 9 constants [25], as opposed to only 6 in eqn (8).

## III. RESULTS

Figure 1 shows the low-temperature solid configuration for simulated monolayers of C8, C10, and C12. The herringbone phase is shown for the solid phase in C8 and C10, and a commensurate RC structure is evident from the solid phase of C12.

Figure 2 contains order parameters $OP_{her}$, $OP_2$, and $OP_4$, where $OP_{her}$ is represented by

$$OPher = \frac{1}{N_m}\left\langle \sum_{i=1}^{N_m} \sin(2\phi_i)(-1)^j \right\rangle, \quad (9)$$

and is a good indicator of herringbone order present within a solid. This order parameter is maximized when $\phi_i=45°$ and $135°$ (with the orientation of the molecules in each sublattice alternating) and takes on a value of unity in such a case. In simulations conducted in this study, $OP_{her}$ takes on values near 0.86 in a HB phase, which corresponds to molecular orientations of about 35° and 150° with respect to the *x*-axis of the cell. These orientations represent a commensurate solid, where the molecules are aligned with the glide-line of the graphite substrate. $OP_{her}$ is used as an indicator of the HB phase for C8 and C10.

The next two order parameters, $OP_2$ and $OP_4$, are special cases of $OP_n$, which is represented by the function

$$OPn = \frac{1}{N_m}\left\langle \sum_{i=1}^{N_m} \cos n(\phi_i) \right\rangle, \tag{10}$$

where *n* takes on values of 2 and 4 for this study. This order parameter is a strong indicator of *n*-fold symmetry that exists in the solid phase (and also, in this case, the intermediate phase). With higher values of *n* used in (10), the corresponding values of $OP_n$ give indications of fluctuations that are present in the molecular orientations of the system. By using values of *n*=2 and *n*=4, the maximum values of $OP_n$ ($OP_n$=1) correspond to those in which the molecules are strictly aligned either normal or parallel to the *x*-axis of the computational cell, depending on the value of *n*. To study the behavior of the intermediate phase, only $OP_2$ is used for C8 and C10, since the intermediate phase is expected to exhibit large thermal fluctuations, and hence, using a higher value of *n* in an order parameter would not be particularly useful. However, for C12, both *n*=2 and *n*=4 are used to study the fluctuations that exist around melting, since the solid phase of C12 is an RC phase. However, to achieve a positive value of $OP_2$ for C12, the absolute value of the order parameter is presented to compare its behavior to $OP_4$.

One more important order parameter that is paramount to classifying commensurate-incommensurate phase transitions is $OP_{com}$, shown in figure 3. This order parameter is defined as

$$OPcom = \frac{1}{3N_m}\sum_{i=1}^{N_m}\left\langle \sum_{s=1}^{6} \exp(-i\vec{g}_s \cdot \vec{r}_i) \right\rangle, \tag{11}$$

and represents the sum of all six reciprocal lattice vectors in graphite for each molecular center-of-mass position. This quantity is maximum (normalized such that $OP_{com}$=1) when all center-of-mass positions are directly over graphite hexagon centers, and becomes vanishingly small when commensurability with the graphite substrate is lost, and the molecular center-of-masses no longer reside over hexagon centers. Another special case (as observed for C8) is when the center-of-mass positions reside directly over maxima on the graphite substrate (i.e. directly between hexagon centers). In such a case, $OP_{com}$ takes on a negative value, and in the case where all molecules are in such positions, $OP_{com}=-1$.

In figures 4-5, a quantitative representation of the presence of gauche defects within the monolayer is presented. Figure 4 represents a distribution that shows the temperature dependence of the end-to-end distance in monolayers of C8, C10, and C12. At low temperatures, this distribution has a single distinct peak corresponding to all molecules being in the trans conformation. As the temperature is increased, more molecules shift into a gauche conformation and multiple peaks corresponding to methyl-methyl distances for the various possible gauche defects arise. This distribution is extremely useful for indicating the type of gauche defects present in the system (centralized or about the endgroups).

In figure 5, another version of the end-to-end distance is presented. In this case, the end-to-end distance for each molecule is averaged over the post-equilibrated period of 500 ps in each simulation, and normalized as to represent the average methyl-methyl distance for the simulated molecules at each temperature point. To a good approximation, in the low-temperature solid phase (well below the transition region), the molecules are in the trans conformation where the methyl-methyl distance is greatest. Therefore, a reference point is chosen 95K below the melting transition and labeled as $L_t$, such that the difference in the average methyl-methyl distance ($L$) at any temperature point with the average methyl-methyl distance in the trans conformation can be analyzed. To compare how the end-to-end distance in the molecules differs for C8, C10, and C12 near the melting transition, the melting transition temperature, as defined by structural order parameters and pair-correlations, is set to be *Tm* for each case and each value of *ΔL* is plotted as a function of *Tm*, where the *x*-axis is shifted such that *Tm* is the same point for each case. A significant value of *ΔL* corresponds to a significant number of molecules in the gauche conformation.

The data in figure 5 can be further analyzed through figure 6, which shows only the intramolecular portion of the atomic pair-correlation function, $g(r)$ for C8, C10, and C12. At low temperatures, there are $n$-1 (where $n$ is the number of simulated pseudoatoms in the molecule) peaks that correspond to intramolecular pair distances in each molecule, however, as the temperature is increased, a series of gauche peaks between these $n$-1 peaks appear. Thus, the intramolecular $g(r)$ gives similar information to figure 5, but better detailed information regarding both the type of gauche defect that is dominant (central or about the endgroups in the molecules), and precise intramolecular distances corresponding to molecules in the gauche conformation.

In figure 7, the temperature dependence of the average atomic height, $P_{avg}(z)$, is shown for all three monolayers studied. This quantity is included to show that there is a sharp increase in $P_{avg}(z)$ for C8 near the intermediate transition, corresponding to significant layer promotion (as observed in simulations of C6 monolayers previously). However, there is no sharp increase in $P_{avg}(z)$ for monolayers of C10 and C12, only a gradual incline that increases consistently through melting. This is indicative that there is *not* a significant amount of layer promotion that occurs at the intermediate phase transition in monolayers of C10 and C12. This is further analyzed through use of snapshots of C8 and C10 monolayers ca. 10K after the intermediate phase transition takes place, shown in figure 8. In both monolayers, there is a degree of nematic-like order present, where C8 seems to best represent the intermediate phase behavior observed in monolayers of C6 previously studied. Comparison of both snapshots indicates that the intermediate phase of C10 exhibits less layer promotion and more molecules exhibiting gauche defects than does C8.

In figure 9, the bond-orientational distribution for 1-5$^{th}$ center-of-mass molecular neighbors is presented. The molecular neighbor shells in this calculation are determined by minima in the intermolecular part of the pair-correlation function (whose intramolecular part is shown in figure 5). To accurately define the neighbor shells, the minima are defined separately for the three separate phases, as to accommodate shifting in intermolecular pair distances due to effects such as layer promotion. The distribution, $P(\phi)$, represents the probability of a neighbor $j$ (within the five neighbor shells) having an orientation $\phi$ relative to a molecule $i$ with respect to the $x$-axis. This quantity has distinct signatures for both a HB phase as well as an RC (or nematic-like) phase due to the unique molecular orientations in each case, and is used to quantify the melting transition, as well as indicate differences in the orientational behavior in each of these phases.

In order to give a good sense for how the melting transition is affected by the two contributing mechanisms, tilting and gauche defects, a series of simulations are carried out for a 100 ps equilibration period and a 200 ps dynamical period by which the dihedral constants in the potential are each increased by 10x, to completely eliminate the formation of gauche defects in the molecules. Shorter simulations are carried out because the time scales over which tilting takes place are significantly less than those that describe how gauche defects form. The transition temperatures are determined by means of order parameters described previously and are reported, along with transition temperatures for all three monolayers studied, in table II. To determine the degree of tilting that the molecules undergo, a tilt order parameter, $OP_{tilt}$, defined as:

$$OP_{tilt} = \frac{1}{2N} \left\langle \sum_{i=1}^{N_m} (3\cos^2 \theta_i - 1) \right\rangle \tag{12}$$

is the thermal average of a Legendere polynomial, with $\theta_i$ representing the orientational out-of-plane angle that the moment of inertia axis of each molecule makes with the substrate. This quantity is used to make statements in the following section about how the tilting behaves when the formation of gauche molecules is eliminated. Since eqn (12) is a measure of the moment of inertia axis, a significant number of gauche defects that are directed out of the surface plane will complicate the averaged value of $OP_{tilt}$ and could indicate a significant amount of tilting, when the large $OP_{tilt}$ value is really due to the formation of gauche defects. Therefore this order parameter is only useful in studying tilting when gauche defects are eliminated, and is used to quantify the tilting behavior in such a case, but is not presented since this quantity indicates absolutely no tilting in monolayers of C10 and C12 in such a case.

Finally, figure 10 presents the difference between the melting temperature when gauche defects are eliminated and the normal simulated monolayer melting temperature ($T_{2(10Xgauche)}$-$T_2$) as a function of alkane chain length. Based upon previous studies over butane monolayers [13] and hexane monolayers [18] where this type of variation was studied, this figure shows the trend of the increasing dependence on the formation of gauche defects from a chain length where only one gauche defect is available (with two possible orientations). In general, Hansen et al. [13,17] report no significant difference with such a variation for butane monolayers, so this difference is fixed at 0K. The author expects that there will be some variation in this temperature difference due to the structural shift that takes place from C10-C12 monolayers, which could be responsible for the slight decrease for C12 in figure 10. However, it is evident that this temperature difference for C12 is comparable to that for C10, so the fit that is presented represents that.

# IV. DISCUSSION

This section will be split into three separate parts. The first part will focus on the phase transition and the possible phase transition mechanism that seems to dominate the commensurate-incommensurate transition (CIT) from the low-temperature solid phase to an intermediate phase. The second part will focus on the observed shifting of the melting transition mechanism, and the effects of alkyl length upon tilting and the formation of gauche defects. The third part will include a short comparison of the UA model to the behavior of the system in simulations that utilize the AUA4 model, and will remark on any significant differences in any of the observed and studied transitions, as well as a brief mention of possible applications of this work.

### A. Intermediate Phase Transition

Previous theoretical work [18] that studies monolayer hexane on graphite finds that there is an intermediate phase that exhibits local nematic order. This corresponds to the experimental observation of a solid/liquid coexistence region, as observed by Newton [12], where the observed behavior from diffraction data seems to indicate the loss of order with respect to the substrate and the formation of rectangular-centered (RC) island patches existing in a liquid. Recent simulations [18,29,30] have pointed out that there seems to be a significant amount of molecular tilting that takes place in concert with some layer promotion at the transition from a commensurate HB solid to an incommensurate nematic-ordered phase. This indicates that the space reduction is facilitated by molecular layer promotion, [29,30] which creates vacancies that could bring about the phase transition. With this in mind, inspection of figure 7 indicates that there are significant differences in the intermediate phases of C8 and C10 as compared to C6 studied previously. First of all, one of the most significant differences is that monolayers of C10 in the intermediate phase show very little or no layer promotion at all. In fact, comparison of the C8 intermediate phase to the C10 intermediate phase shows that there are quite a few more C10 molecules that exhibit various gauche defects about their molecule chains. These snapshots alone already suggest that there could be a significant difference in how the intermediate phase takes place and behaves for C8, C10, and C12 monolayers as compared to C6.

This is further evidenced through data for the intermediate phase transition, $T_1$, and the end-to-end distance difference between the trans conformation in table II and figure 5 respectively. From table II, $T_2$-$T_1$ (difference between the intermediate and melting phase transition temperatures) for C8 is ca. 40K. Comparing this to figure 5, 40K below the melting temperature there is very little change in methyl-methyl distance from the trans conformation, indicating that there is some formation of gauche defects that contribute to a space reduction at the intermediate phase transition, but not very significant as compared to C10 and C12. In fact, from table II, $T_2$-$T_1$ for C10 is ca. 30K and for C12 is ca. 20K, which corresponds to points in figure 4 that indicate quite a bit of difference in average methyl-methyl distance as compared to the trans conformation. This can only attributed to a more significant number of gauche molecules at the intermediate phase transition, which in turn contributes to the vacancy formation needed for the system to undergo the phase transition into an intermediate phase. From this data, there is an indication that an increased number of gauche defects near the transition in C10 and C12 could be responsible for a greater space reduction, and hence explain the decrease in layer promotion shown in figure 7. In fact, this is further shown in figure 8 from the temperature dependence of the average atomic height, $P_{avg}(z)$. For C8 monolayers, it is evident that at the intermediate phase transition ($T$=188K) there is a very sharp increase in the average atomic height that takes place between the temperatures of $T$=185K and 188K. This very sharp increase is not observed for monolayers of C10 and C12, but rather a gradual increase is observed, which could indicate a larger total number of gauche defects present in the monolayer. In most cases, it is energetically favorable for the gauche defect to protrude out of the surface plane in the solid phase, therefore, an increased number of gauche defects would increase the average atomic height as well. Such a gradual increase is most likely a signature of the increasing number of gauche molecules with an increasing thermal energy in the system.

Further evidence of the evolution of gauche defects before and during the intermediate phase is shown in figure 6. For monolayers of C8, the intramolecular $g(r)$ indicates only 7 distinct intramolecular neighbor peaks in the solid HB phase, with only a *very* small gauche peak present between the $2^{nd}$ and $3^{rd}$ intramolecular neighbors in the solid phase at $T$=180K (ca. 8K before the intermediate phase transition). However, in comparison to monolayers of C10, temperature points in the HB phase ca. 25K *below* the intermediate phase transition (well in the solid HB phase) already indicate a presence of gauche defects that is even more significant than any presence observed in C8 monolayers with temperatures ca. 17K *nearer* to the intermediate phase transition. This emphasizes the difference that exists in monolayers of C8 and C10 regarding how gauche defects form in the molecules prior to the intermediate phase transition, and how much the intermediate phase transition is affected by the increased planar space that is created by these molecular conformation changes.

Finally, through a series of additional simulations that are described in section 3, the dihedral potential constants are increased by an order of magnitude in order to eliminate the formation of gauche defects. More than anything, this process is a way of studying how the out-of-plane tilting mechanism contributes to the phase transitions, and in this case, the intermediate phase transition. Although various tilt order parameters can be formulated to study the tilting, the out-of-plane effects of the gauche molecules makes it impossible to differentiate tilting from a perspective of the moment of inertia, used in such order parameters. Therefore, table II indicates that all systems studied have a dependence on the formation of gauche defects to undergo the phase transitions. However, for this portion of the study, these additional simulations indicate that there is absolutely *no* tilting that takes place for monolayers of C10 leading up to the intermediate phase transition, but there is a *very* small amount of tilting (probably insignificant as compared to that needed for a "space reduction") that takes place prior to the C8 intermediate phase transition (and hence, the transition takes place at much lower temperatures). Interestingly, the C10 monolayer does not indicate any layer promotion, even when gauche defects are completely eliminated, whereas the C8 monolayer indicates quite a few molecules promoted to the 2$^{nd}$ layer in such a case. This could suggest that it is energetically unfavorable for the molecules to promote to the 2$^{nd}$ layer due solely to the molecular chain length. Also, in both cases, the intermediate phase transition takes place at a significantly higher temperature with the formation of gauche defects eliminated than without. This indicates that the intermediate phase is dependent upon the space reduction created by gauche defects, but that the effects on the intermediate phase transition (as well as the melting transition, as will be discussed next) are much more dramatic in the case of C10, indicating that C10 is heavily, if not *solely* dependent upon gauche defects to carry out this transition.

One should take the results of this section with an important caveat. First of all, the "intermediate" phase transition for C10 and C12 give completely different behaviors. In C8 and C10 monolayers, the intermediate phase transition proceeds similar to that observed in previous work over C6 monolayers, with the transition into an intermediate (incommensurate) phase with RC order from the solid HB phase. However, the "intermediate" phase for C12 monolayers refers to the loss of strict RC order in the solid monolayer, which exists between the well-defined solid phase and the disordered phase. Therefore, one must differentiate these two types of behaviors as both are refereed to the "solid-intermediate phase transition."

**B. Melting Phase Transition**

There has been previous work that studies the phase transitions in monolayers of butane and hexane previously, [17] and these studies come to the final conclusion that the phase transition mechanism relies almost solely upon tilting in butane, and mostly upon tilting in hexane. [16-18] Recent work [18] has found that there is an existent dependence upon gauche defects in hexane monolayers (i.e. by eliminating gauche defects, the melting transition temperature was found to rise by ca. 20K), but that tilting is the dominant effect that contributes to melting.

First of all, from analysis of $OP_2$ and $OP_4$ in figure 2, intermolecular $g(r)$'s (not shown), and 1-5$^{th}$ bond-orientational distributions in figure 9, this study finds that the melting temperatures of C8, C10, and C12 are $T_2$=225±2K, 255±2K, and 285±3K respectively. Directly comparing the simulated melting temperature for C12 to the solid monolayer melting temperature observed through calorimetry measurements [4] ($T$=287.8K) shows that both are in excellent agreement within the simulated uncertainties that arise from having to increment simulation temperatures. Further, although no calorimetry data is completed for monolayers of C8 and C10 that indicate precise solid monolayer melting temperatures, Castro et al. [5] do, in fact, find that the solid monolayer melts at a higher temperature than the bulk melting point. Without considering uncertainties, the simulated monolayer temperature for C8 and C10 are ca. 8K and 12K respectively above the bulk melting point for each case, which seems to be consistent with the experimental observations that this work reports. Therefore, in terms of melting temperatures, the simulation results seem to be in good agreement with what has been previously observed experimentally for these systems.

In terms of the melting transition, the effects of gauche defects and tilting for C8, C10, and C12 are much different than those observed for C4 and C6 monolayers in previous work. This is especially evident in figure 6. Where previous work over C6 [18] observes a very small number of gauche defects forming in the "nematic" phase, this work observes a large amount of gauche defects forming in the intermediate/nematic phase for both C8 and C10 monolayers which is evident in figure 6 from the significant gauche peaks that exist between the 2-4$^{th}$ intramolecular neighbors. Figure 6 also indicates that the gauche peaks are quite a bit larger in both the intermediate as well as the fluid phases for C10 as compared to C8. This indicates a more significant presence of molecules exhibiting gauche defects, which further emphasizes the effects of the chain length upon the formation of gauche defects in the monolayer.

In figure 4 it is evident that even in the intermediate phase, there is a significant presence of gauche defects about the endgroups of the molecules in all cases of C8, C10, and C12. However, in the case of C10 and C12, the presence of *centralized* gauche defects becomes much more significant as the temperature is increased to that near the melting temperature of the monolayer. In such cases, the amplitude of the distribution peak representing molecules in the trans conformation nearly disappears, and is replaced by a series of peaks corresponding to molecules exhibiting gauche defects about their ends as well as their central regions. Although C8 shows some indication of this behavior, it is much more apparent for C10 and C12 monolayers. Another intriguing aspect of figure 4 is the peak that exists near and after the melting temperature that has a very small ($\Delta L$~3.5Å) methyl-methyl separation in monolayers of C10 and C12. This is interpreted as chain-melting of the molecules themselves, where centralized gauche defects contribute to a loss of intramolecular order within the molecules. This type of behavior is exhibited in longer *n*-alkanes [1,2] as well, and is in some sense similar to behavior near the well-known gel-to-fluid transition in lipid bilayers that longer chain alkanes are found to represent well.

The methyl-methyl distance at the melting temperature for C8, C10, and C12 as compared to the trans methyl-methyl distance in figure 5 shows that $\Delta L$ for C10 at the melting temperature is nearly *three times* that of $\Delta L$ for C8 at melting. Further, $\Delta L$ for C12 is quite a bit larger than that of C10 as well. This underlines the effects of gauche defects at the melting transition, and how these affect each monolayer. Additionally, the data for melting in C8, C10, and C12 with 10x dihedral constants indicates that there is ca. 50K difference in the melting temperature for C8 monolayers, but for C10 and C12 monolayers, there are 130K and 115K differences between the melting temperature with 10x dihedral constants and the normal melting temperature. This is visually evident from figure 10, where these differences are plotted as a function of the chain length, with results from previous studies over C4 and C6 monolayers included. From this figure, one can observe a trend that the melting transition has a nonlinear dependence on the formation of gauche defects as a function of chain length until a chain length near that of C10, where there is a change in the behavior of this trend. Although it is unclear how this trend evolves as the chain length is increased even further, the similar behaviors of C10 and C12 monolayers infers a different type of chain length dependence on gauche defects (which is assumed in figure 10 to be generally constant) than the smaller chained alkanes. One interpretation of this behavior (that which is offered in this study) is that C10 and C12 are completely reliant upon gauche defect formation to undergo the phase transition, whereas smaller chained alkanes involve other space reduction mechanisms, such as layer promotion and/or tilting in order to proceed with the phase transition. If such is the case, then C10 plays a significant role as the first molecule in the family of linear *n*-alkanes to have the necessary chain length in order for the internal degrees of freedom to determine the melting behavior.

### C. AUA Model and Relevant Applications

Temperature sequences utilizing the AUA4 model are used for two primary reasons which are (*i*) to further validate the UA model and the behavior observed (the UA model is found to be in good agreement with both the AUA model and with experiment in previous work over hexane), since the UA model distinguishes each molecule only by mass and (*ii*) to involve a second perspective of how the system melts through use of second form of a dihedral potential. Although one would expect the effects of anisotropy in the methyl groups to have less effect as the length of the molecule increases, this second model provides a check to the model that is used to report the results found in sections A and B, which depend strongly on the modeled intramolecular behavior.

The general phase behavior observed with the AUA model is very similar to that observed with the UA model. One of the most significant differences in the phases is the presence of more HB order in the intermediate phase of C8 and C10 than is observed for the UA model, as well as minimal layer promotion in the intermediate phase of C8. Another significant effect that comes about with the AUA model is that there are significantly less molecules that are in a gauche conformation in the intermediate phases. However, by performing short simulations of 10x dihedral constants for the AUA model, the results indicate that the melting transition, instead of being increased by ca. 50K for C8 with the UA model, is increased by ca. 70K with the AUA model. This could be due to the small overestimation in the moment of inertia that the UA model has been found exhibit previously, which indicates that the melting transition in C8 monolayers is a bit more dependent upon gauche defects, but there seems to be no evidence that this is a sole dependence. In all cases, the melting transitions exhibited no larger than a 10K difference between those observed for the UA model, so this additional study concludes that the UA model and the AUA model are in general agreement, with some small different aspects (such was reported in a previous work over monolayer C6 as well).

The short-chained *n*-alkanes are well known for their excellent lubrication properties in many practical applications. In fact, monolayers of alkanes are especially important for their contribution to thin lubricating films at the interface of two solid surfaces. [31] Such films tend to decrease the friction and wear that arises from the rigidity of the two solids, and allow for greater longevity of the solid materials. In relation to this particular study, the author speculates that the dynamics of lubrication in thin films that are dominated by molecules in a gauche conformation would be particularly different than the case of shorter-chained alkanes, where the molecules tend to be more rigidly in trans conformation. In fact, in such a case (at the solid-alkane-solid interface), the effects of confinement of the alkane monolayer by the solid surfaces would be expected to play a role in the solid-fluid phase transition in itself, since melting occurs largely based upon out-of-plane behavior for these monolayers. Therefore, it could potentially useful to understand the differences in how the melting transition takes place (since confinement could actually have less effect on those monolayers that disorder via gauche defects as opposed to those that disorder via rigid molecular out-of-plane tilting) in order to better understand how to possibly manipulate these systems to give a desired type of behavior.

## V. CONCLUSIONS

The study at hand involves a couple of key concepts that extends an understanding of melting in the studied monolayers. First of all, the presence of gauche defects before the melting transition in C10 and C12 monolayers significantly exceeds that exhibited by the C8 monolayers. Variations indicate that C10 monolayers seem to be the first to break the nonlinear dependence of the melting transition vs. alkane chain length, and C12 monolayers seem to be comparably dependent upon gauche defects as compared to C10 monolayers. This seems to indicate that C10 is the first linear chain molecule in the *n*-alkane family whose monolayer melting behavior is dominated by the internal degrees of freedom. Secondly, the loss of the solid-phase order in C8, C10, and C12 monolayers seem to be facilitated less by layer promotion as the chain length of the alkane monolayer increases. Based upon the increased presence of gauche defects prior to the CIT in C10 and C12, this behavior is largely attributed to the greater presence of gauche defects which tend to create a lower density cell and satisfy the needed space reduction for the phase transition (although, this may not be so much the case in the C12 monolayers due to the solid phase order). Therefore, this study shows that gauche defects play a major role in how the melting behavior takes place as the chain length of the molecules adsorbed onto the graphite substrate increases, with monolayers with molecules having a chain length involving 10 carbon atoms being the first to exhibit melting behavior completely dependent upon the molecular degrees of freedom.

## ACKNOWLEDGEMENTS

The author greatly acknowledges Paul Gray and the UNI CNS for use of the UNI Opteron cluster for calculations. Also, special thanks to the UNI Physics Dept. for use of computing facilities, as well as John Deisz for assistance with the operation of these facilities.

# TABLES AND FIGURES

| Type | $a$ (Å) | $b$ (Å) | Cell size (Å) | Temp. range (K) | $N_m$ | $N_a$ |
|---|---|---|---|---|---|---|
| Octane (C8) | 4.9 | 21.2 | 85.217 x 68.88 | 120-250 | 112 | 896 |
| Decane (C10) | 5.1 | 25.5 | 76.695 x 81.6 | 140-290 | 96 | 960 |
| Dodecane (C12) | 4.26 | 35.0 | 76.68 x 70 | 190-300 | 72 | 864 |

**Table I.** Unit cell parameters, computational cell sizes, simulated temperature ranges, and the total number of molecules and psuedoatoms in each simulation for monolayers of C8, C10, and C12. Unit cell parameters are used from ref. [7].

|  | Octane (C8) | Decane (C10) | Dodecane (C12) |
|---|---|---|---|
| $T_1$ | 188±2K | 225±2K | 265±3K |
| $T_2$ | 225±2K | 255±2K | 285±3K |
| 10x dihedral $T_1$ | 255±5K | 365±5K | N/A |
| 10x dihedral $T_2$ | 275±5K | 385±5K | 400±10K |

**Table II.** Phase transition temperatures $T_1$ and $T_2$ for monolayers of C8, C10, and C12, as well as in the case where the constants in the dihedral potential are increased by an order of magnitude. The use of N/A for C12 indicates that, with the precision used to simulate temperature points in this region, the transition into an intermediate phase was not observable.

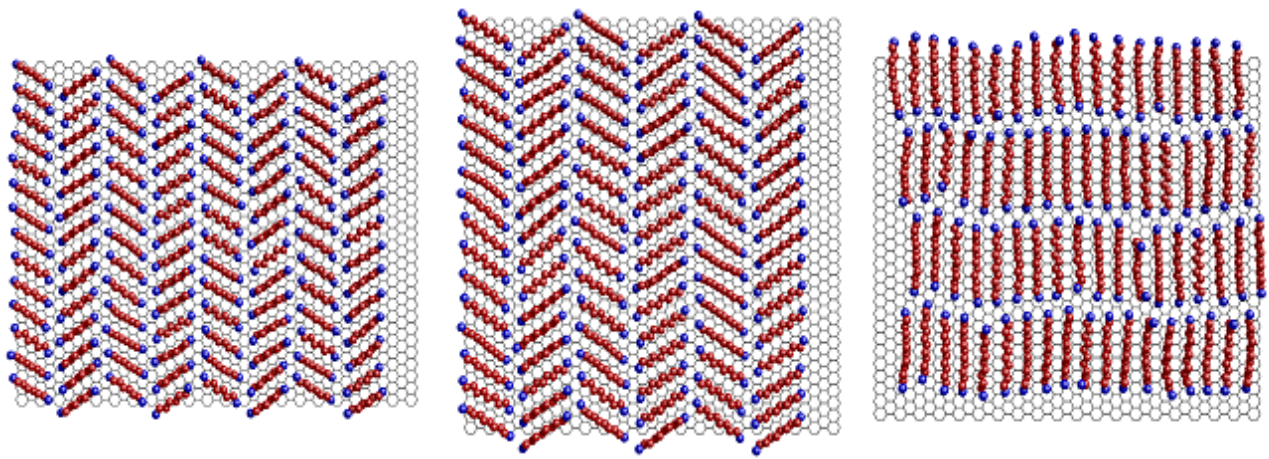

(online color) **Figure 1.** Snapshots of the low-temperature solid phases exhibited by monolayers of C8, C10, and C12 respectively after 700 ps at $T$=120K (C8), $T$= 140K (C10), and $T$=230K (C12).

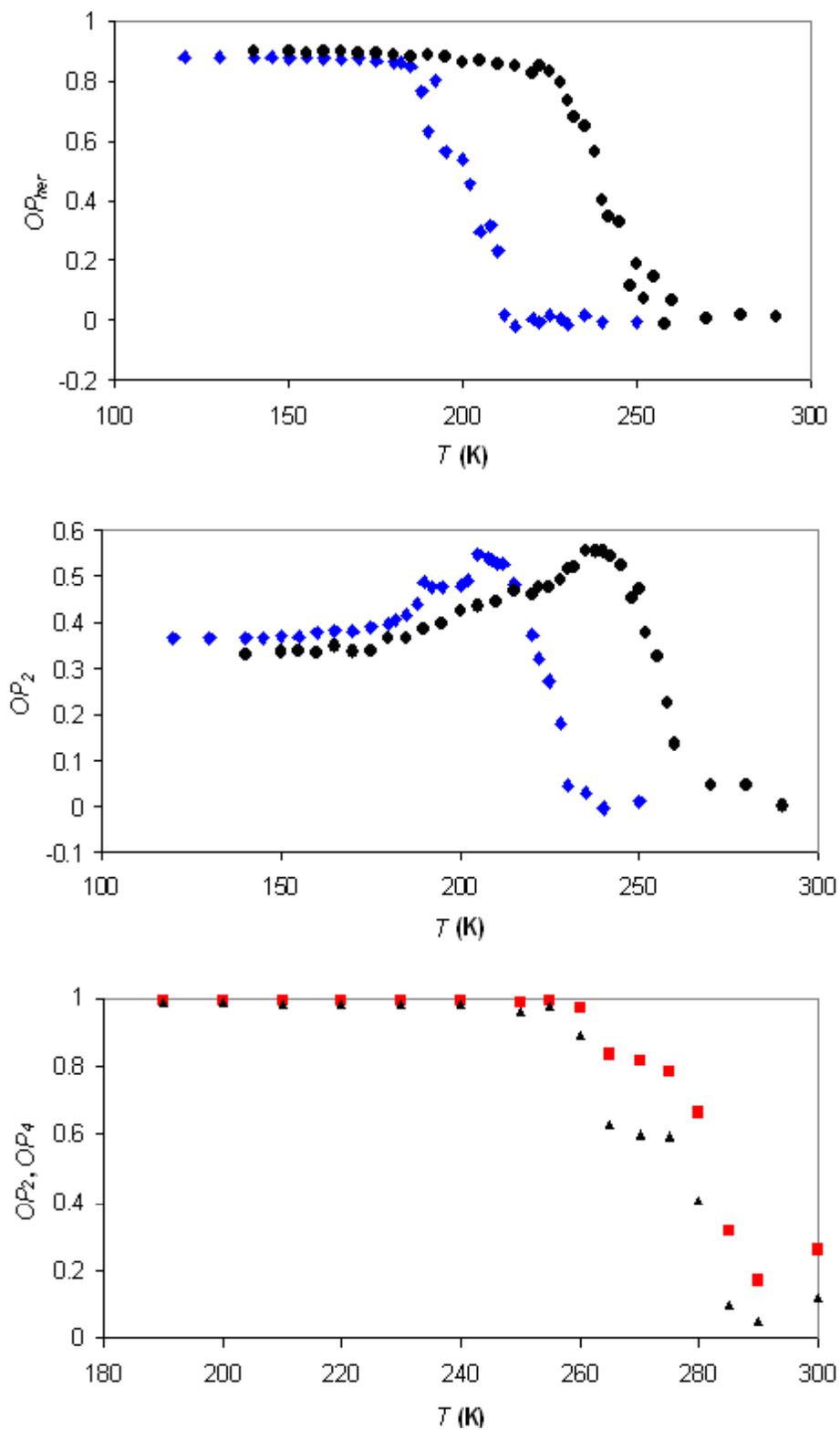

(online color) **Figure 2.** (top 2) $OP_{her}$ and $OP_2$ for C8 (diamonds) and C10 (circles) monolayers, and (bottom) $OP_2$ (squares) and $OP_4$ (triangles) for C12 monolayers.

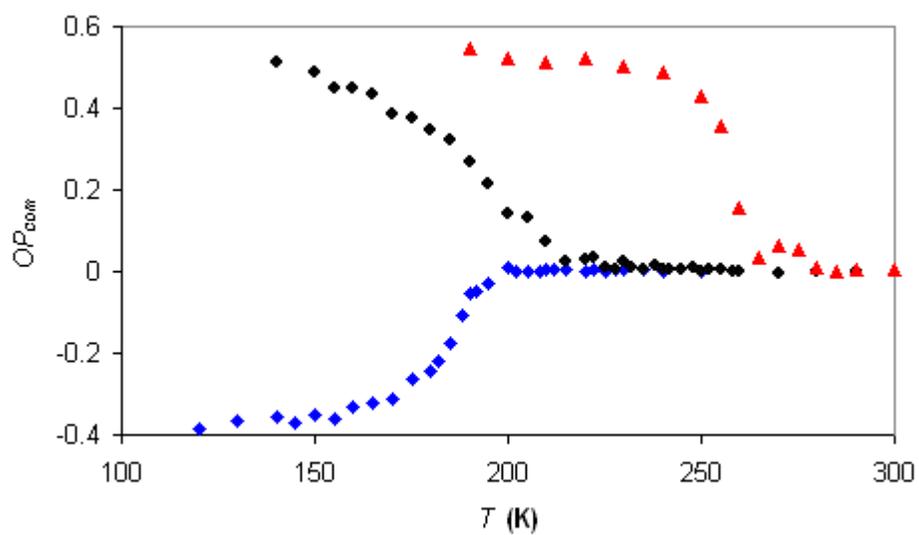

(online color) **Figure 3.** Temperature dependence of $OP_{com}$ for monolayers of C8 (diamonds), C10 (circles), and C12 (triangles) respectively.

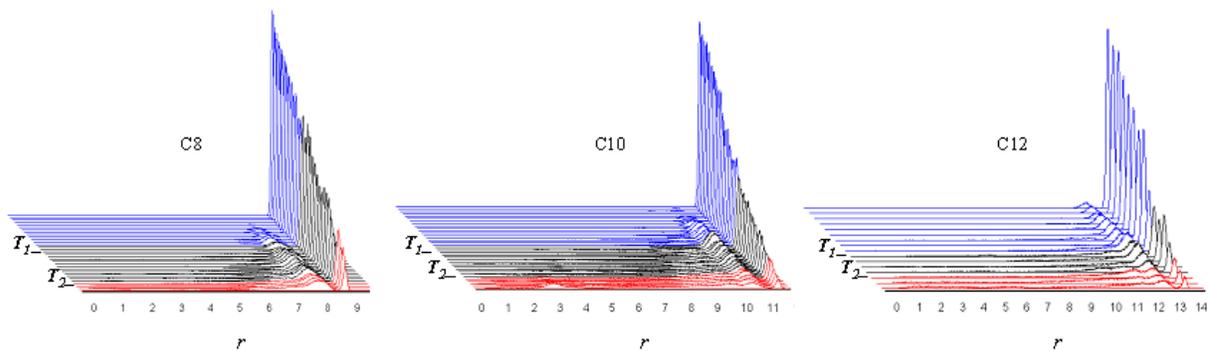

(online color) **Figure 4.** End-to-end distributions in Angstroms for monolayers of C8, C10, and C12 respectively. The temperature increases from back to front, with the plotted distributions (blue) behind $T_1$ corresponding to the solid HB phase, the distributions between $T_1$ and $T_2$ (black) corresponding to the intermediate phase, and the distributions in front of $T_2$ (red) corresponding to the disordered fluid phase.

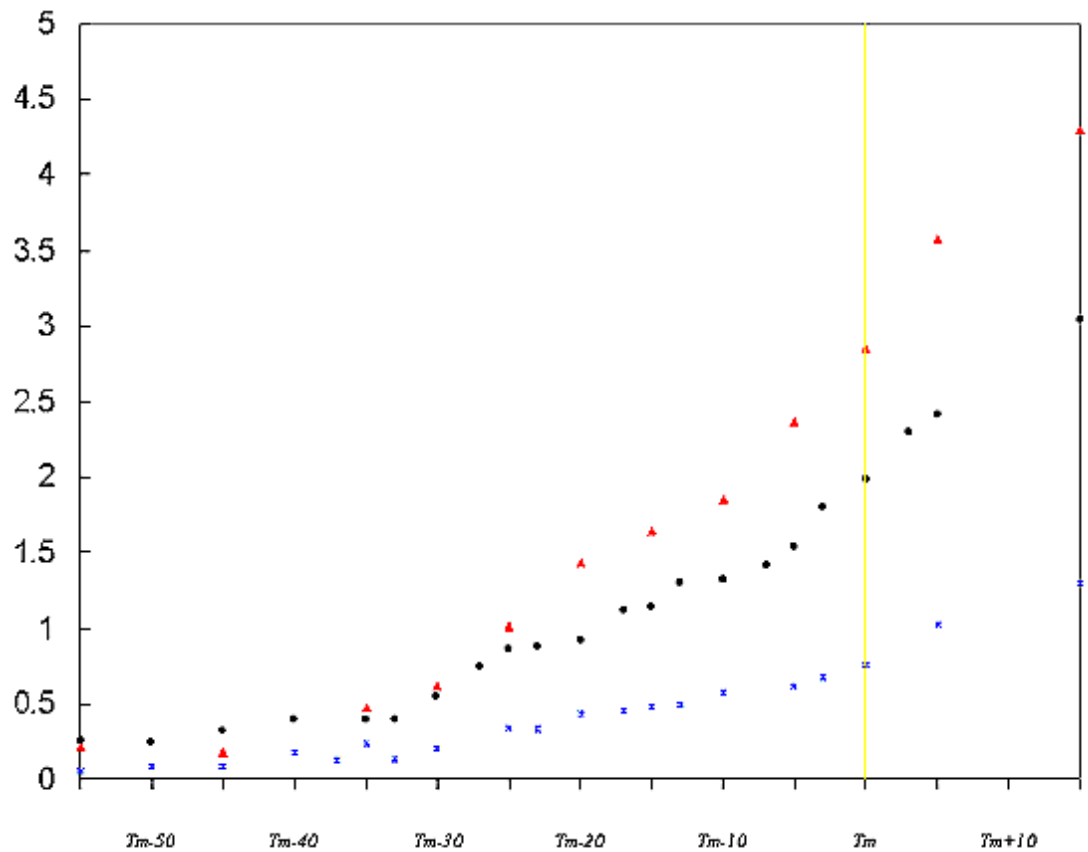

(online color) **Figure 5.** Difference in the methyl-methyl distance and the trans methyl-methyl distance in Angstroms as a function of the melting temperature (*Tm*). The triangles represent monolayer C12, the circles represent C10, and the stars C8. The line represents the melting temperature for each case.

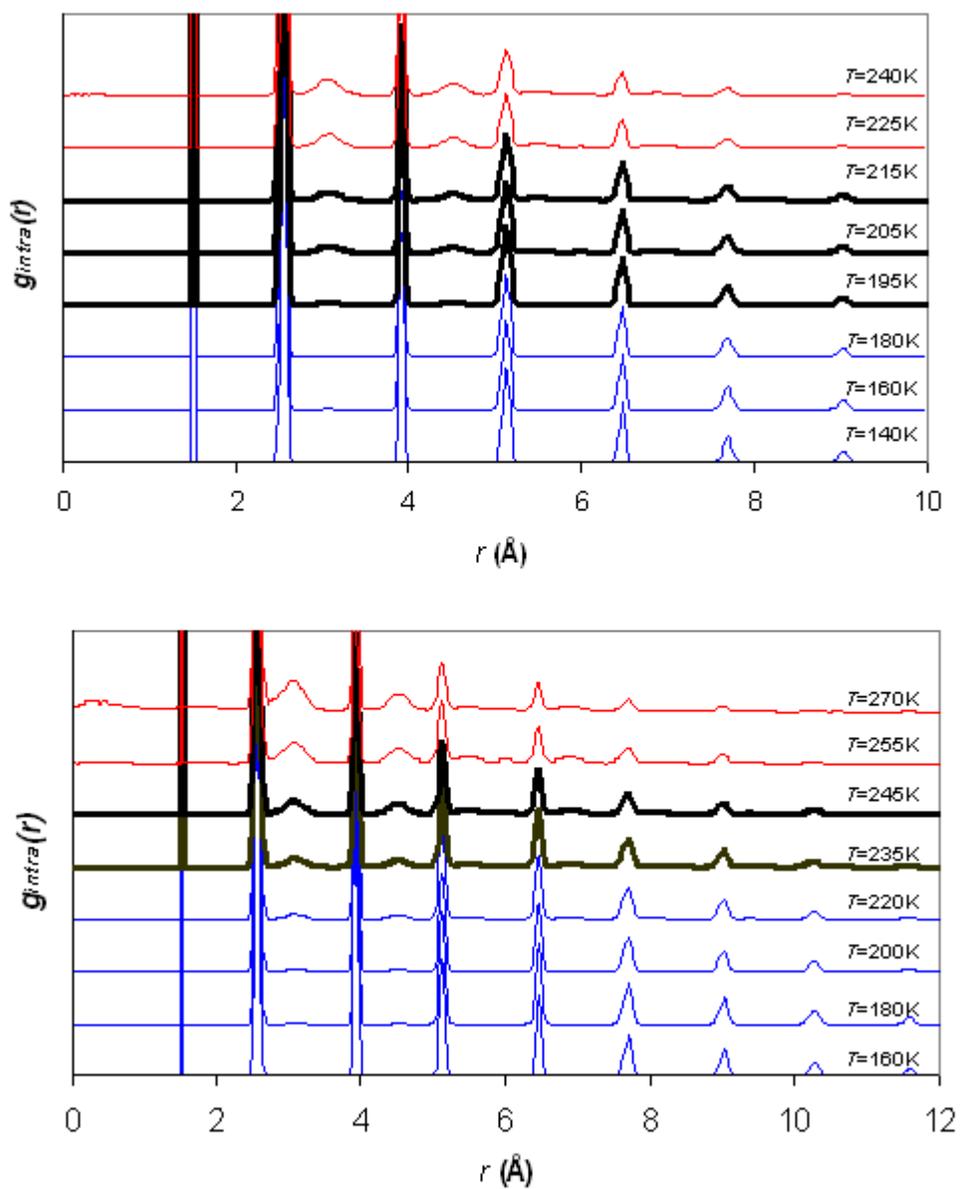

(online color) **Figure 6.** Intramolecular *g(r)* for monolayers of C8 (top) and C10 (bottom). The points in the solid phase are on bottom (blue), the points in the intermediate phase are in the middle in bold (black), and the points in the fluid are on top (red).

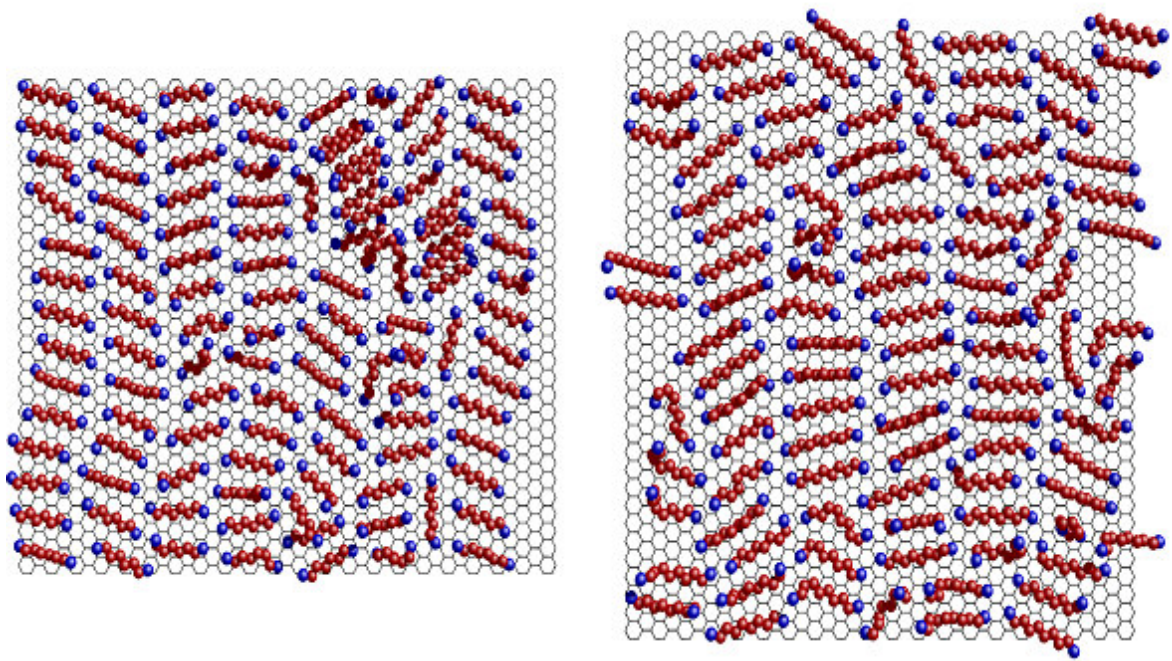

(online color) **Figure 7.** Snapshots of the intermediate phases in both C8 and C10 monolayers (ca. 10K above $T_1$). Notice how the C8 monolayer (left) seems to display both nematic order and layer promotion, whereas the C10 monolayer (right) indicates a more significant presence of molecules with gauche defects and no layer promotion.

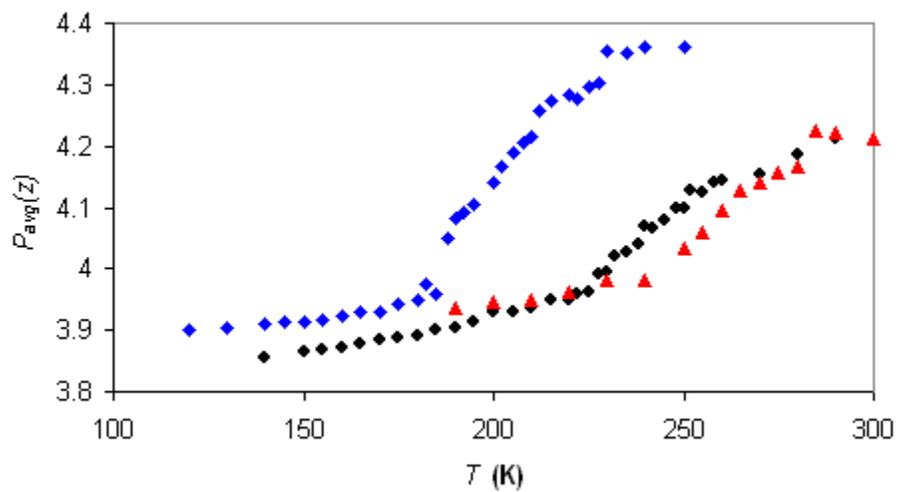

(online color) **Figure 8.** Average *P(z)* for monolayers of C8 (diamonds), C10 (circles), and C12 (triangles) at simulated temperature points.

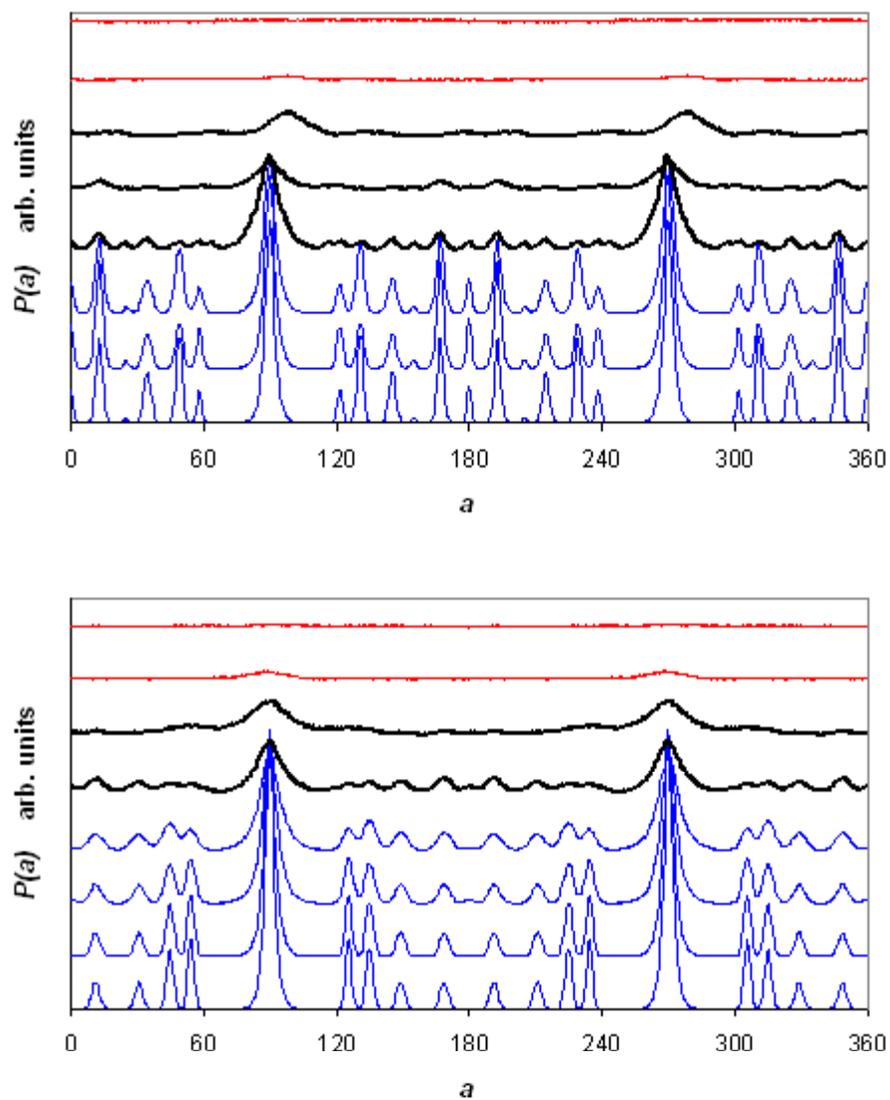

(online color) **Figure 9.** 1-5$^{th}$ Bond Orientational distributions, *P(a)*, for monolayers of C8 (top) and C10 (bottom). The labeling convention and temperature points are identical with those in figure 5.

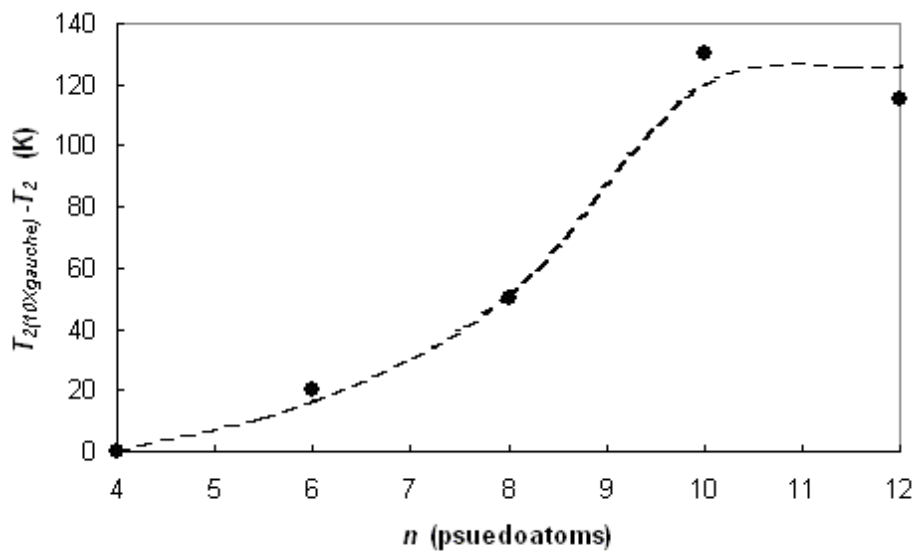

**Figure 10.** Difference in the melting temperature without the formation of gauche defects and the melting temperature of the simulated solid monolayer as a function of the chain length (in psuedoatoms). The $n=4$ and $n=6$ points are taken from [13, 17] and [18] respectively. The dotted line gives the proposed fit to these five points.